\documentclass{aastex63}

\received{}
\revised{}
\accepted{}
\submitjournal{ApJ}

\shorttitle{ALMA observations of NGC~604}
\shortauthors{Muraoka et al.}

\begin{document}

\title{ALMA Observations of Giant Molecular Clouds in M33. II. Triggered High-mass Star Formation by Multiple Gas Colliding Events at the NGC~604 Complex}

\correspondingauthor{Kazuyuki Muraoka}
\email{kmuraoka@p.s.osakafu-u.ac.jp}

\author[0000-0002-3373-6538]{Kazuyuki Muraoka}
\affil{Department of Physical Science, Graduate School of Science, Osaka Prefecture University, 1-1 Gakuen-cho, Naka-ku, Sakai, Osaka 599-8531, Japan}

\author{Hiroshi Kondo}
\affil{Department of Physical Science, Graduate School of Science, Osaka Prefecture University, 1-1 Gakuen-cho, Naka-ku, Sakai, Osaka 599-8531, Japan}

\author[0000-0002-2062-1600]{Kazuki Tokuda}
\affiliation{Department of Physical Science, Graduate School of Science, Osaka Prefecture University, 1-1 Gakuen-cho, Naka-ku, Sakai, Osaka 599-8531, Japan}
\affiliation{National Astronomical Observatory of Japan, National Institutes of Natural Science, 2-21-1 Osawa, Mitaka, Tokyo 181-8588, Japan}

\author[0000-0003-0732-2937]{Atsushi Nishimura}
\affiliation{Department of Physical Science, Graduate School of Science, Osaka Prefecture University, 1-1 Gakuen-cho, Naka-ku, Sakai, Osaka 599-8531, Japan}

\author[0000-0001-8187-7856]{Rie E. Miura}
\affiliation{National Astronomical Observatory of Japan, National Institutes of Natural Science, 2-21-1 Osawa, Mitaka, Tokyo 181-8588, Japan}

\author{Sachiko Onodera}
\affiliation{Meisei University, 2-1-1 Hodokubo, Hino, Tokyo 191-0042, Japan}

\author{Nario Kuno}
\affiliation{Department of Physics, Graduate School of Pure and Applied Sciences, University of Tsukuba, 1-1-1 Tennodai, Tsukuba, Ibaraki 305-8577, Japan}
\affiliation{Tomonaga Center for the History of the Universe, University of Tsukuba, Tsukuba, Ibaraki 305-8571, Japan}

\author[0000-0001-6149-1278]{Sarolta Zahorecz}
\affiliation{Department of Physical Science, Graduate School of Science, Osaka Prefecture University, 1-1 Gakuen-cho, Naka-ku, Sakai, Osaka 599-8531, Japan}
\affiliation{National Astronomical Observatory of Japan, National Institutes of Natural Science, 2-21-1 Osawa, Mitaka, Tokyo 181-8588, Japan}

\author[0000-0002-2794-4840]{Kisetsu Tsuge}
\affiliation{Department of Physics, Nagoya University, Chikusa-ku, Nagoya 464-8602, Japan}

\author[0000-0003-2062-5692]{Hidetoshi Sano}
\affiliation{National Astronomical Observatory of Japan, National Institutes of Natural Science, 2-21-1 Osawa, Mitaka, Tokyo 181-8588, Japan}

\author{Shinji Fujita}
\affiliation{Department of Physical Science, Graduate School of Science, Osaka Prefecture University, 1-1 Gakuen-cho, Naka-ku, Sakai, Osaka 599-8531, Japan}

\author[0000-0001-7826-3837]{Toshikazu Onishi}
\affiliation{Department of Physical Science, Graduate School of Science, Osaka Prefecture University, 1-1 Gakuen-cho, Naka-ku, Sakai, Osaka 599-8531, Japan}

\author[0000-0003-1549-6435]{Kazuya Saigo}
\affiliation{National Astronomical Observatory of Japan, National Institutes of Natural Science, 2-21-1 Osawa, Mitaka, Tokyo 181-8588, Japan}

\author[0000-0002-1411-5410]{Kengo Tachihara}
\affiliation{Department of Physics, Nagoya University, Chikusa-ku, Nagoya 464-8602, Japan}

\author{Yasuo Fukui}
\affiliation{Department of Physics, Nagoya University, Chikusa-ku, Nagoya 464-8602, Japan}
\affiliation{Institute for Advanced Research, Nagoya University, Furo-cho, Chikusa-ku, Nagoya 464-8601, Japan}

\author[0000-0001-7813-0380]{Akiko Kawamura}
\affiliation{National Astronomical Observatory of Japan, National Institutes of Natural Science, 2-21-1 Osawa, Mitaka, Tokyo 181-8588, Japan}



\begin{abstract}

We present the results of ALMA observations in $^{12}$CO($J=2-1$), $^{13}$CO($J=2-1$), and C$^{18}$O($J=2-1$) lines and 1.3\,mm continuum emission
toward a massive ($\sim 10^6$\,$M_{\odot}$) giant molecular cloud associated with the giant H$\,${\sc ii} region NGC~604
in one of the nearest spiral galaxy M33 at an angular resolution of 0\farcs44 $\times$ 0\farcs27 (1.8\,pc $\times$ 1.1\,pc).
The $^{12}$CO and $^{13}$CO images show highly complicated molecular structures composed of a lot of filaments and shells whose lengths are 5~--~20\,pc.
We found three 1.3\,mm continuum sources as dense clumps at edges of two shells and also at an intersection of several filaments.
We examined the velocity structures of $^{12}$CO($J=2-1$) emission in the shells and filaments containing dense clumps, and concluded that expansion of the H$\,${\sc ii} regions cannot explain the formation of such dense cores.
Alternatively, we suggest that cloud--cloud collisions induced by an external H$\,${\sc i} gas flow and the galactic rotation compressed the molecular material into dense filaments/shells as ongoing high-mass star formation sites.
We propose that multiple gas converging/colliding events with a velocity of a few tens km\,s$^{-1}$ are necessary to build up NGC~604, the most significant cluster-forming complex in the Local Group of galaxies.

\end{abstract}

\keywords{stars: formation --- ISM: clouds--- ISM:  kinematics and dynamics --- ISM: individual (NGC~604) --- galaxies: Local Group}

\section{Introduction} \label{sec:intro}

Stars are formed by the contraction of the molecular interstellar medium (ISM), which would be initiated by spontaneous gravitational collapse or some triggering events, such as cloud--cloud collisions (converging flow), galactic shocks, and feedback from high-mass stars \citep[e.g.,][]{mckee2007,dobbs2014}.
In the past few decades, many observational studies toward the Local Group of galaxies (see the reviews by e.g., \citealt{blitz2007,fukui2010,heyer2015}) suggested that giant molecular clouds (GMCs) are major sites of high-mas star formation,
and the formed stars eventually regulate the evolution of galaxies through their feedbacks such as radiation pressure, photoionization, and supernova explosion \citep{hopkins2012}.
Thus, the detailed study of GMC properties and the evolution of GMCs is one of the most important issues in modern astronomy.

Earlier molecular-gas surveys using CO molecule and its isotopes toward star forming regions in the Milky Way (MW) revealed that
filamentary structures are considered to be fundamental ingredients of molecular clouds \citep[e.g.,][]{mizuno1995,onishi1996,goldsmith2008,hacar2013}, and such filamentary molecular clouds eventually collapse into dense cores prior to star formation. In addition, high-spatial resolution observations of molecular clouds outside the MW with ALMA are now providing new insights;
recent ALMA observations toward the Large Magellanic Cloud (LMC) suggest that the collision/interaction of the filamentary molecular clouds may drive the high-mass star formation.
For example, \cite{fukui2015} found that the CO distribution at a subparsec scale is highly elongated with a small width at the active star-forming region N159 West in the LMC.
These elongated clouds (i.e., filaments) show straight or curved shape with a typical width of 0.5~--~1.0\,pc and a length of 5~--~10\,pc.
They also detected the molecular outflow toward the high-mass protostar located at the intersection of two spatially overlapping filaments,
and thus they argued that the two filaments collided with each other and triggered the formation of high-mass stars (see also \citealt{saigo2017} for N159 East).

The role of galactic dynamics is also considered to be important for the onset of high-mass star formation.
In disks of spiral galaxies, molecular clouds and star forming regions are preferentially distributed in spiral arms.
This is because molecular gas can be accumulated in the potential minimum of stellar arms,
and then the formation of dense molecular gas and high-mass stars are triggered
due to galactic shock \citep{fujimoto1968,roberts1969,shu1973} caused by quasi-stationary density waves \citep[e.g.,][]{lin1964} and/or due to cloud--cloud collisions.
Thus, it is significant to investigate the relation between galactic-scale gas dynamics and internal structures of GMCs for the further understanding of the high-mass star formation.

The flocculent spiral galaxy M33 is one of the most preferable objects for this purpose.
Its proximity ($\sim$840\,kpc; \citealt{freedman2001}) and favorable inclination (51$^{\circ}$; \citealt{corbelli2000})
enable us to resolve internal molecular structures of individual GMCs within the M33 disk by using ALMA.
So far, a lot of observational studies on ISM and star formation has been performed toward M33 \citep[e.g.,][]{engargiola2003,heyer2004,rosolowsky2007,onodera2010,gratier2010,tosaki2011,miura2014,druard2014,corbelli2017,sano2020}, and individual GMCs have been identified \citep{gratier2012,miura2012,corbelli2017}.
In particular, a massive ($\sim 10^6$\,$M_{\odot}$) GMC associated with NGC~604, which exists in the spiral arm (see figure~\ref{fig:IRAMHST}) and is the second-most luminous supergiant H$\,${\sc ii} region in the Local Group after 30~Dor in the LMC, has been studied in various wavelengths;
radio \citep{viallefond1992,wilson1992,wilson1997,churchwell1999,tosaki2007,miura2010,muraoka2012,tachihara2018}, optical \citep{drissen1993,yang1996,hunter1996,tenorio2000,gonzalez2000,bruhweiler2003,maiz2004,relano2009}, and infrared \citep{eldridge2011,farina2012,martinez2012}.
According to these earlier studies, NGC~604 contains more than 200 young (3~--~5 Myr) O-type stars associated with a bright H$\alpha$ nebula,
which seems to be composed of multiple shells, extending up to a radius of 200~--~400\,pc.
In addition, the effect of the stellar feedback on the neighboring molecular gas is often discussed.
Considering that a huge amount of molecular gas is still remaining in the close vicinity of NGC~604, which is different from the condition of 30~Dor,
NGC~604 is a unique target to investigate both the triggering mechanism of star formation and the feedback from young clusters.

In this paper, we present the results of ALMA observations with an angular resolution of 0\farcs44 $\times$ 0\farcs27
(corresponding to 1.8\,pc $\times$ 1.1\,pc at the distance of M33) toward the GMC associated with NGC~604 (hereafter, NGC~604-GMC).
We also performed such high-angular resolution ALMA observations toward two other GMCs in M33, GMC-16 and GMC-8 catalogued by \cite{miura2012},
which are in different evolutionary stages from the NGC~604-GMC.
The detailed results of GMC-16 are already reported by \cite{tokuda2020} and those of GMC-8 will be reported in a forthcoming paper (H. Kondo et al.\,2020, in preparation).

The structure of this paper is as follows.
In Section~\ref{sec:obs}, we describe the ALMA observations and data reduction.
Then, we reveal the internal molecular structure of the GMC and its kinematics based on the moment maps of CO($J=2-1$) lines,
and also present the new finding of 1.3\,mm continuum sources as sites of ongoing star formation in Section~\ref{sec:results}.
In Section~\ref{sec:discuss}, we firstly discuss the possible triggering mechanism of high-mass star formation in the NGC~604-GMC,
and then, we investigate the formation processes of the observed molecular structures and those of dense clumps and high-mass stars at the 1.3\,mm continuum sources.
In particular, we discuss the role of the galactic-scale gas dynamics in high-mass star formation within parsec-scale molecular clouds.

\section{Observations and Data Reduction} \label{sec:obs}

\begin{figure*}[ht!]
\epsscale{1.15}
\plotone{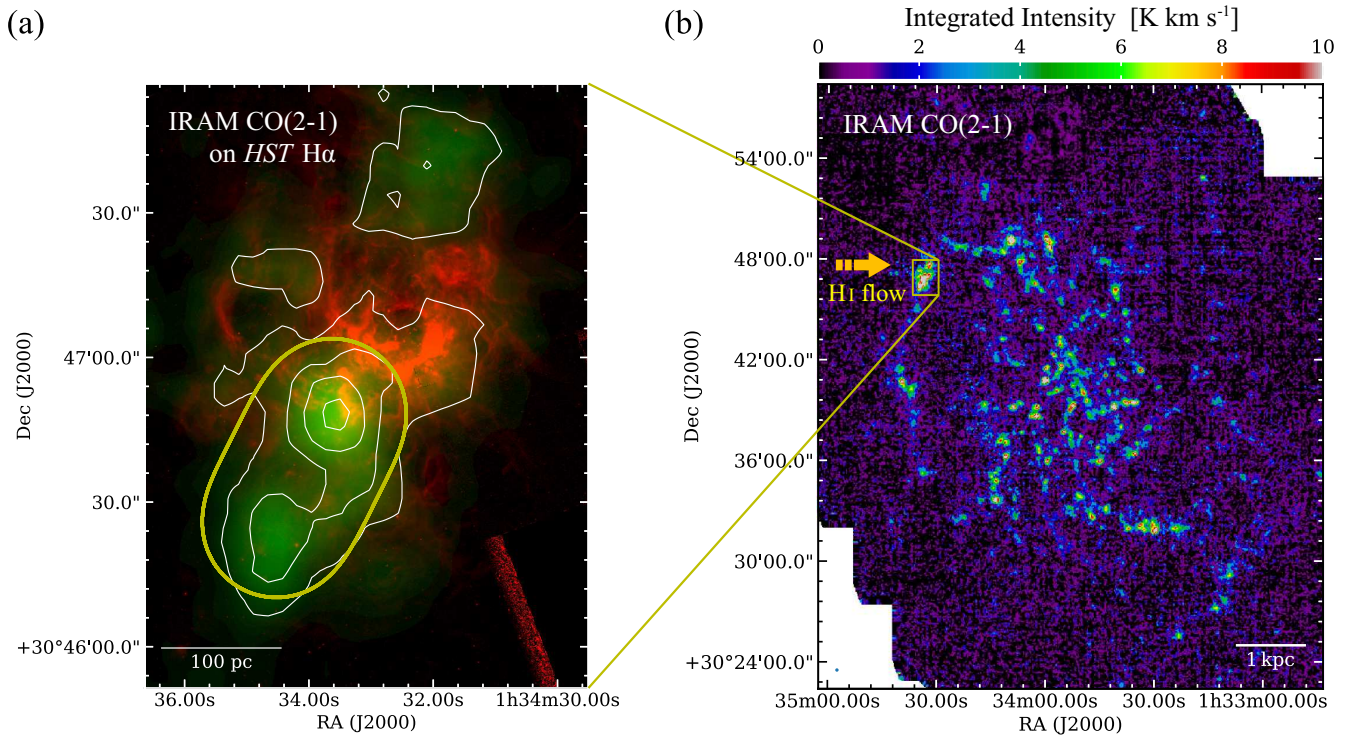}
\caption{(a) Integrated intensity map in $^{12}$CO($J=2-1$) emission of NGC~604 obtained by the IRAM 30-m telescope (white contour and green; \citealt{druard2014})
superposed on the spatial distribution of H$\alpha$ emission obtained with the $HST$ (red), which is available in the $HST$ archive.
The contour levels are 3, 6, 9, and 12\,K\,km\,s$^{-1}$.
The yellow line indicates the field of view of the ALMA observations.
(b) Integrated intensity map in $^{12}$CO($J=2-1$) emission of M33 obtained by the IRAM 30-m telescope \citep{druard2014}.
The angular resolution is shown in the lower left corner.
The arrow indicates the infalling H$\,${\sc i} gas flow toward NGC~604 proposed by \cite{tachihara2018} (see subsection~\ref{res:CO}).
}\label{fig:IRAMHST}
\end{figure*}

ALMA Cycle 6 observations toward NGC~604-GMC were carried out in Band~6 (211~--~275\,GHz)
with the main array 12\,m antennas and the Atacama Compact Array (ACA) 7\,m antennas between 2017 October and 2018 October.
The target field was composed of three adjacent pointings, which could cover most of the main GMC in NGC~604, as shown in figure~\ref{fig:IRAMHST}(a).

The target molecular lines were $^{12}$CO($J=2-1$), $^{13}$CO($J=2-1$), and C$^{18}$O($J=2-1$).
The bandwidths of the correlator settings were 117.19\,MHz with 1920\,channels for the $^{12}$CO line and 960\,channels for $^{13}$CO and C$^{18}$O lines.
In addition, we used two spectral windows for the 1.3\,mm continuum observations with an aggregate bandwidth of 3750\,MHz with a channel width of 0.98\,MHz.

We briefly summarize the imaging parameters and resultant data qualities. The detailed processes of data reduction are the same as \cite{tokuda2020}.
We used the \texttt{tclean} task with \texttt{multi-scale} deconvolver \citep{kepley2020}, which is imprinted in the Common Astronomy Software Application (CASA) package \citep{mcmullin2007}.
The weighting scheme of \texttt{tclean} was $``$Briggs$"$ with a robust parameter of 0.5.
We combined the 12\,m and 7\,m array data with the \texttt{feathering} task.
The beam size of CO data is 0\farcs44 $\times$ 0\farcs27, and the rms noise level is $\sim$1.1\,K at a velocity resolution of $\sim$0.2\,km\,s$^{-1}$.
The beam size and the sensitivity of 1.3\,mm continuum are 0\farcs41 $\times$ 0\farcs25 and 0.019\,mJy beam$^{-1}$, respectively.

We evaluated the missing flux of the $^{12}$CO emission by comparing the single-dish data obtained by the IRAM 30-m telescope \citep{druard2014}.
The flux loss is not significant, not more than $\sim$30\% in the 12\,m + 7\,m array data at the northern CO peak (figure~\ref{fig:IRAMHST}),
whereas the southern side has a $\sim$50\% missing flux. We thus combined the ALMA and the IRAM data using the \texttt{feathering} technique.
Note that the ALMA program did not include the Total Power (TP) array observations.
We consider that the 12\,m + 7\,m data of $^{13}$CO, C$^{18}$O, and 1.3\,mm have no significant missing flux
because the spatial distributions of such dense gas tracers are intrinsically more compact than that of $^{12}$CO.

\section{Results} \label{sec:results}

\subsection{Molecular Gas Properties in $^{12}$CO and $^{13}$CO} \label{res:CO}

\begin{figure*}[ht!]
\epsscale{1.15}
\plotone{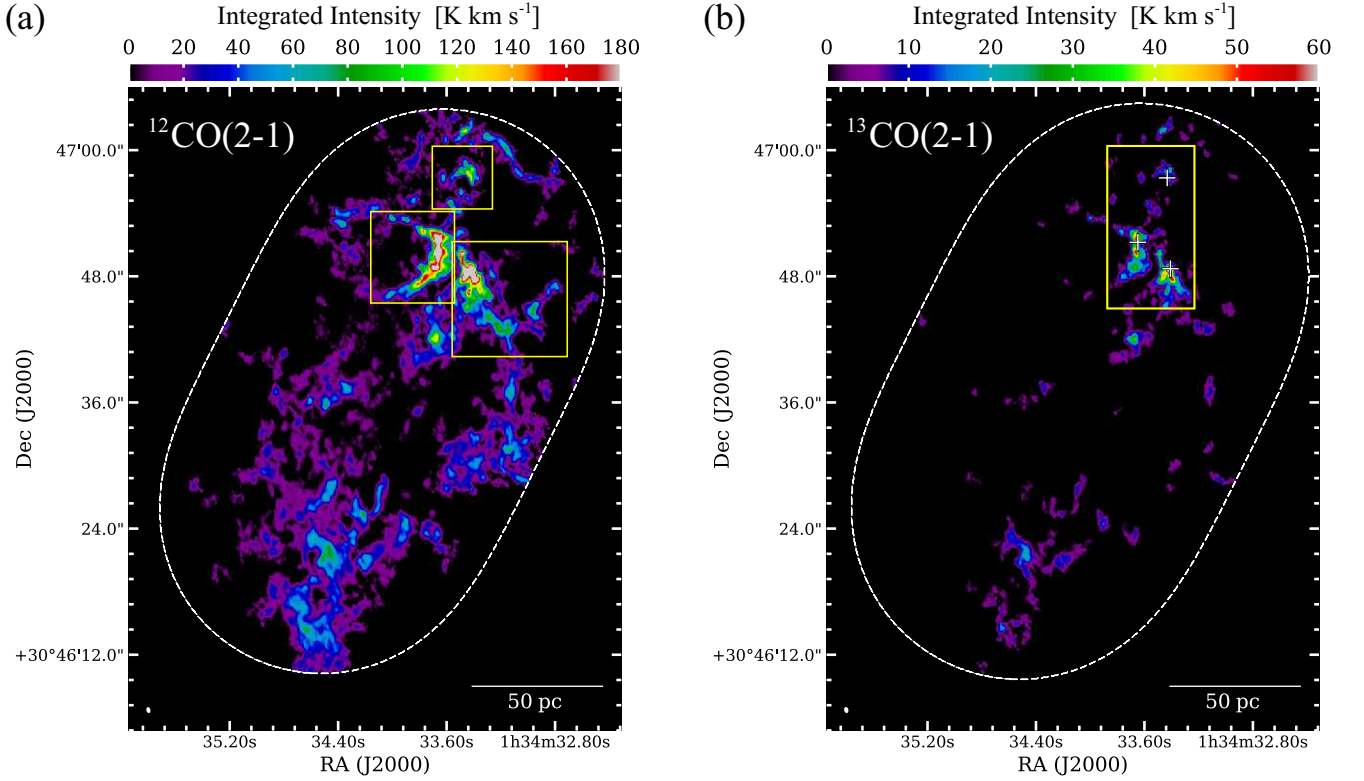}
\caption{(a) Integrated intensity map in $^{12}$CO($J=2-1$) emission of NGC~604.
The dashed white line indicates the field of view of the ALMA observations.
The three rectangles indicate individual molecular shells which are magnified and shown in figure~\ref{fig:12COMMS}.
The synthesized beam is shown in the lower left corner.
(b) Integrated intensity map in $^{13}$CO($J=2-1$) emission.
The rectangle indicates the high molecular-gas density region where 1.3\,mm continuum emission (crosses) is detected (see figure~\ref{fig:MMS}). 
}
\label{fig:12CO13CO}
\end{figure*}

Figure~\ref{fig:12CO13CO} shows 0th moment (i.e., integrated intensity) maps in $^{12}$CO($J=2-1$) and $^{13}$CO($J=2-1$) emission.
Although the previous single-dish studies only revealed the two local peaks of the GMC (see figure~\ref{fig:IRAMHST}(a)), the new $^{12}$CO($J=2-1$) map clearly depicts complicated molecular-gas structures over the observed field.
This is the first parsec-scale GMC view at the vicinity of a supergiant H$\,${\sc ii} region in external galaxy disks. We can see several shell- or arc-like CO structures in the northern field,
and there exists a lot of molecular filaments with a length of 5~--~20\,pc elongated toward various directions and small (typically less than 10\,pc) clumps in the southern field.
The $^{13}$CO($J=2-1$) map shows less extended distributions compared with that of $^{12}$CO, demonstrating that this molecule traces moderately dense gas whose density is $\gtrsim$10$^{3}$\,cm$^{-3}$.
The strong $^{13}$CO($J=2-1$) emission is mainly observed in the northern field, in particular, along molecular shells seen in $^{12}$CO.

We estimated the H$_2$ gas mass traced by each molecule.
We applied the $X_{\rm CO}$ factor, 4\,$\times$10$^{20}$\,cm$^{-2}$\,(K\,km\,s$^{-1}$)$^{-1}$ \citep{druard2014}, and a CO($J=2-1$)/CO($J=1-0$) intensity ratio of 0.85 toward H$\,${\sc ii} regions (the ratio in the Orion-KL region reported by \cite{nishimura2015}).
The total gas mass traced by the $^{12}$CO line ($M_{\rm ^{12}CO}$) is $\sim$3\,$\times$10$^6$\,$M_{\odot}$, which is consistent with the single-dish measurements \citep{gratier2012}.
The local thermodynamic equilibrium (LTE) approximation was applied to estimate the total mass from the $^{13}$CO observations ($M_{\rm ^{13}CO}$)
using the excitation temperature of $^{12}$CO derived from the peak brightness temperature of the $^{12}$CO line and assuming the relative molecular abundance ([H$_2$]/[$^{13}$CO]) of 1.4\,$\times$10$^{6}$ (see the justification by \citealt{tokuda2020}).
The resultant $M_{\rm ^{13}CO}$ is calculated to be $\sim$3\,$\times$10$^{5}$\,$M_{\odot}$, and thus the mass ratio ($M_{\rm ^{13}CO}$/$M_{\rm ^{12}CO}$) is $\sim$10\%.

\begin{figure*}[ht!]
\epsscale{1.15}
\plotone{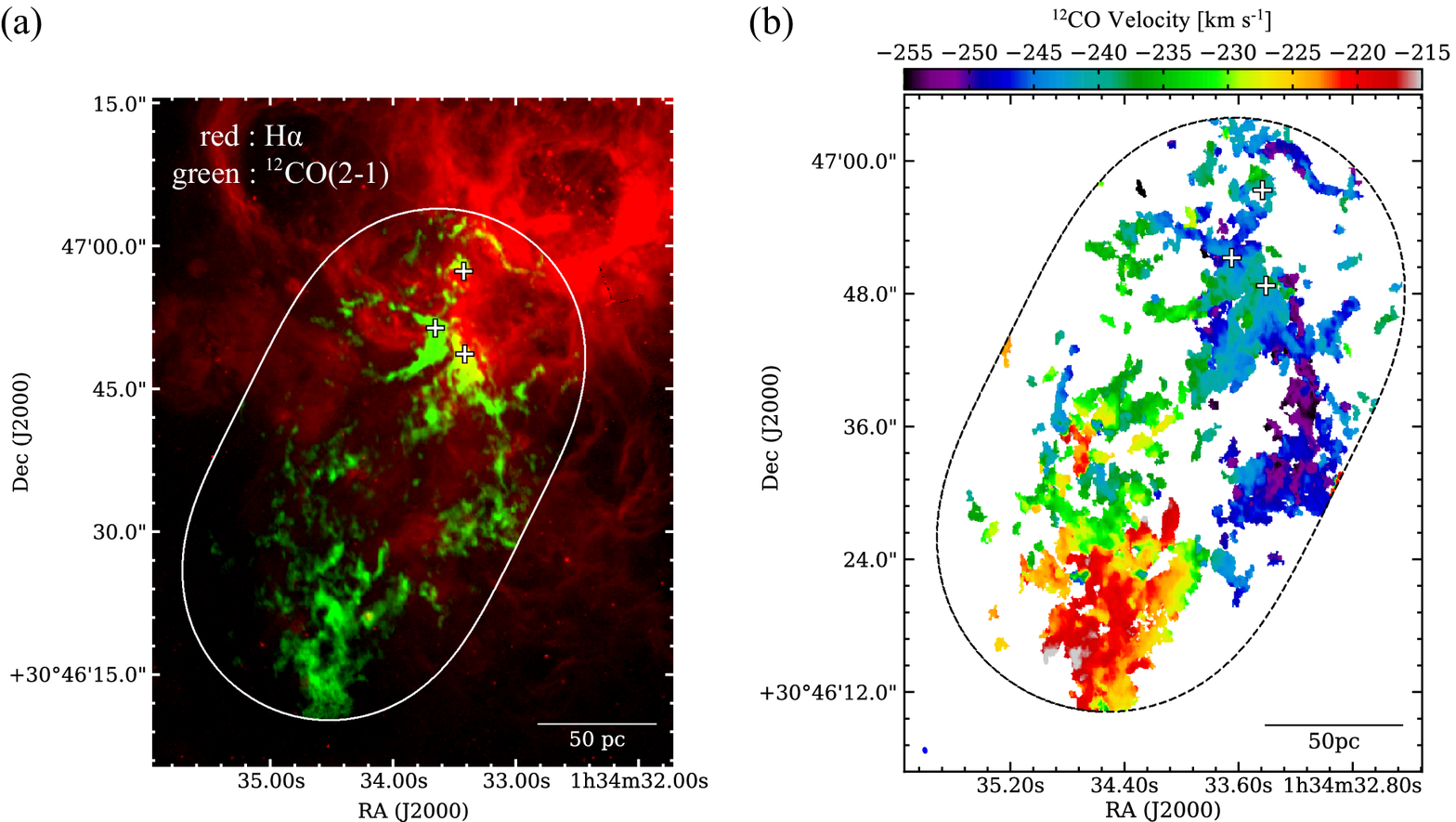}
\caption{(a) Two-color composite image (green: $^{12}$CO($J=2-1$) intensity by ALMA, red: H$\alpha$ by $HST$) of NGC~604.
The white line indicates the field of view of the ALMA observations.
The crosses are the same as those in figure~\ref{fig:12CO13CO}(b).
(b) Velocity field measured in $^{12}$CO($J=2-1$) emission of NGC~604.
The synthesized beam is shown in the lower left corner.
}\label{fig:12COmom1}
\end{figure*}

Figure~\ref{fig:12COmom1}(a) shows two-color maps of the ALMA $^{12}$CO($J=2-1$) integrated intensity (green) and the H$\alpha$ luminosity (red) obtained with the Wide Field and Planetary Camera 2 on the Hubble Space Telescope ($HST$), which clearly demonstrates that some molecular shells are associated with H$\,${\sc ii} regions.
This suggests that the H$\,${\sc ii} regions affect the neighboring molecular gas;
in other words, the feedback should be considered to discuss the structures and physical properties of the molecular shells.
We found that the peak temperature of $^{12}$CO($J=2-1$) emission exceeds 40\,K at the CO peak in molecular shells.
This is presumably due to the heating by UV radiation from high-mass stars within the H$\,${\sc ii} regions.

Figure~\ref{fig:12COmom1}(b) shows the 1st moment map (i.e., velocity field) in $^{12}$CO($J=2-1$) emission over the observed field.
We can see the difference in the representative CO velocities between northern side ($-250$~--~$-240$\,km\,s$^{-1}$) and southern side ($-230$~--~$-220$\,km\,s$^{-1}$) in the GMC.
Such a velocity difference has been already reported by earlier studies \citep[e.g.,][]{miura2010,muraoka2012,druard2014}.
In particular, \cite{tachihara2018} provided an important clue to understand this velocity difference;
they found that the H$\,${\sc i} clouds in NGC~604 are composed of two velocity components separated by 20\,km\,s$^{-1}$.
The red component corresponds to the ``ordinary'' velocity expected from the overall rotation of the M33 disk, while the blue component is considered to be infalling H$\,${\sc i} gas from the eastern side of the GMC (see figure~\ref{fig:IRAMHST}(b)).
Note that \cite{tachihara2018} used the velocity-shifted H$\,${\sc i} spectra in their analyses in order to compensate the velocity gradient due to the inclination and galaxy rotation of M33,
and thus the two H$\,${\sc i} velocity components are centered at $-185$\,km\,s$^{-1}$ and $-165$\,km\,s$^{-1}$, respectively.
We confirmed that the original H$\,${\sc i} velocity components in NGC~604 are centered at $\sim -250$\,km\,s$^{-1}$ and $\sim -230$\,km\,s$^{-1}$, which are almost coincident with the two velocity components observed in our ALMA CO data.
Since the representative CO velocity in the northern side of NGC~604 corresponds to the blue component, the infalling H$\,${\sc i} gas likely affects the properties of the molecular gas and the onset of star formation in NGC~604.
This will be discussed in Section~\ref{sec:discuss}.

\subsection{1.3\,mm Continuum and C$^{18}$O emission} \label{res:1.3mm}

\begin{figure*}[ht!]
\epsscale{0.5}
\plotone{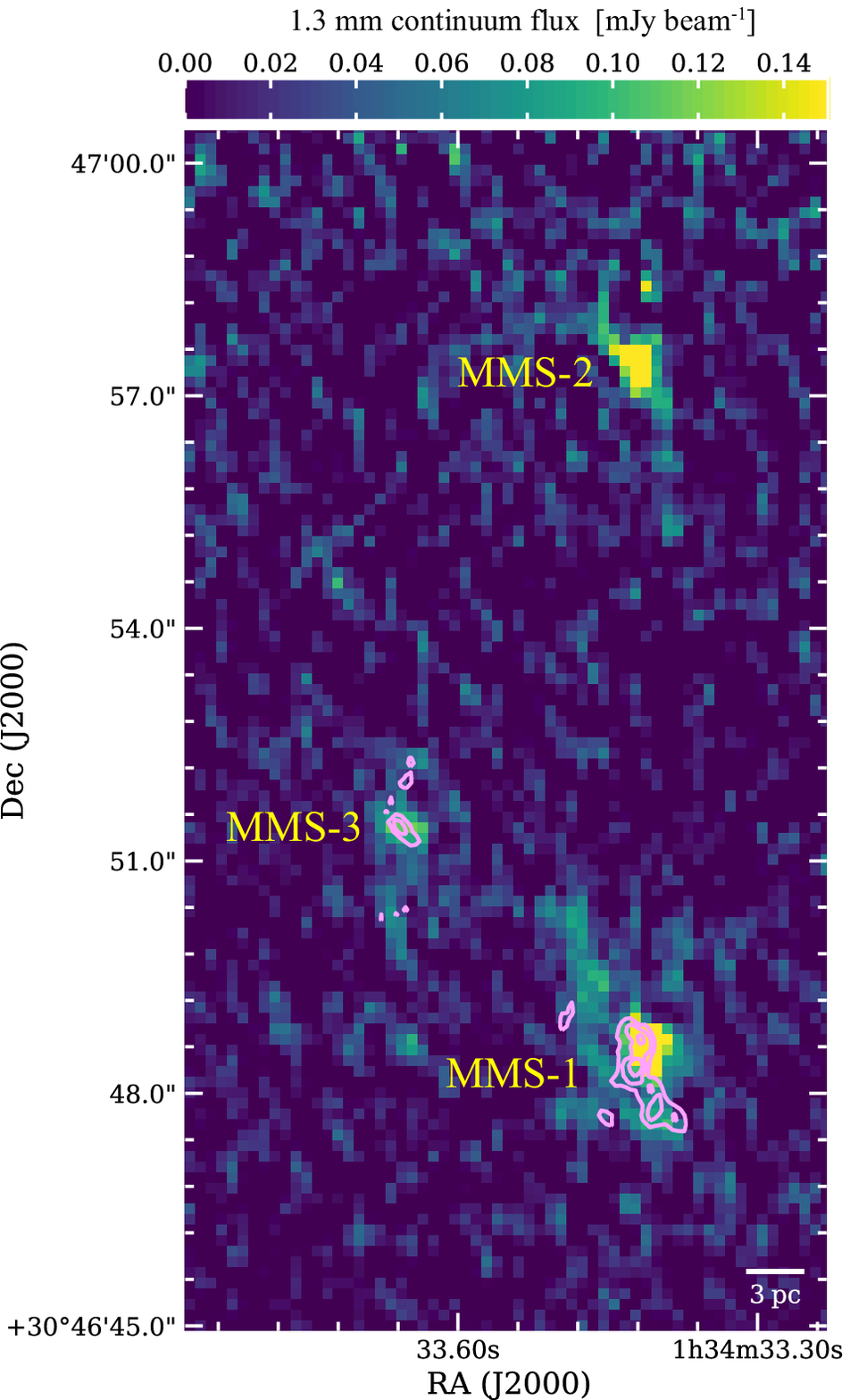}
\caption{Maps of 1.3\,mm continuum flux (color) and the integrated intensity in C$^{18}$O($J=2-1$) emission (contour) of NGC~604.
The contour levels are 3, 6, and 9$\sigma$, where 1$\sigma$ = 0.53\,K\,km\,s$^{-1}$.
Three continuum sources are significantly detected.
}\label{fig:MMS}
\end{figure*}

We found three 1.3\,mm continuum sources (MMSs) in the northern part of the observed field as shown in figure~\ref{fig:MMS}.
The ID number of the MMSs is assigned according to the 1.3\,mm continuum flux, MMS-1, MMS-2, and MMS-3 (see table~\ref{tab:MMS}).
The three sources have bright mid-infrared (MIR) emission and they were categorized as massive young stellar object candidates whose enclosed stellar mass is a few $\times$10$^3$\,$M_{\odot}$ \citep{martinez2012}.
MMS-1, MMS-2, and MMS-3 correspond to Src~1, Src~4, and Src~2 in their catalog, respectively.
In addition, two of the three MMSs have counterparts at centimeter wavelength.
\cite{churchwell1999} reported the 8.4\,GHz (3.6\,cm) continuum map of NGC~604 at an angular resolution of 2\arcsec (8\,pc), and found six 8.4\,GHz components.
Their components A and B are coincident with the positions of our MMS-2 and MMS-1, respectively.
Since the 8.4\,GHz continuum corresponds to the free-free emission from ionized gas, which implies the existence of high-mass stars,
the observed MMSs in NGC~604 are considered to be high-mass star forming regions embedded in dust.
Note that the counterpart of MMS-3 is not identified as an individual component in the 8.4\,GHz continuum map.
This is presumably due to the insufficient angular resolution of the 8.4\,GHz map.

We found difference in the detection of C$^{18}$O($J=2-1$) emission among these MMSs. Remarkable C$^{18}$O($J=2-1$) emission with peak temperature of 1\,K is observed at MMS-1, whereas we found no C$^{18}$O($J=2-1$) emission at MMS-2.
We detected moderate C$^{18}$O($J=2-1$) emission whose peak temperature is 0.6\,K at MMS-3. Considering that the MMSs are sites of ongoing high-mass star formation,
such a difference in the detection of C$^{18}$O($J=2-1$) emission likely reflects different evolutionary stages of star formation (see subsection~\ref{dis:MMS}).

\begin{deluxetable*}{cccccccccc}
\tabletypesize{\scriptsize}
\tablewidth{0pt} 
\tablenum{1}
\tablecaption{Properties of 1.3\,mm continuum sources
\label{tab:MMS}}
\tablehead{
\colhead{Source ID} & \colhead{R.A.} & \colhead{Decl.} & \colhead{Size} & \colhead{$F_{\rm peak}$} & \colhead{$F_{\nu}$} & \colhead{$\Delta V$} & \colhead{$M_{\rm vir}$} & \colhead{$n$(H$_{2}$)} & \colhead{$M_{\rm total}^{MMS}$} \\[-1.0mm]
                    &                &                 & (parsec)       & (mJy\,beam$^{-1}$)       & (mJy)               & (km\,s$^{-1}$)           & ($M_{\odot}$)           & (cm$^{-3}$)            & ($M_{\odot}$) \\[-1.0mm]
\colhead{(1)}       & \colhead{(2)}  & \colhead{(3)}   & \colhead{(4)}  & \colhead{(5)}            & \colhead{(6)}       & \colhead{(7)}            & \colhead{(8)}           & \colhead{(9)}          & \colhead{(10)}
}
\startdata 
MMS-1 & 01 34 33.42 & 30 46 48.59 & 2.9 $\times$ 1.9 & 0.37 & 1.2 & 6.0 & 1.8\,$\times$10$^4$ & 1.8\,$\times$10$^4$ & \textbf{4.5\,$\times$10$^4$} \\
MMS-2 & \textbf{01 34 33.42} & \textbf{30 46 57.31}& 3.2 $\times$ 1.6 & 0.29 & 0.9 & 6.0 & 1.7\,$\times$10$^4$ & 1.9\,$\times$10$^4$ & \textbf{3.3\,$\times$10$^4$} \\
MMS-3 & \textbf{01 34 33.63} & \textbf{30 46 51.38} & 3.2 $\times$ 2.2 & 0.12 & 0.5 & 5.7 & 1.8\,$\times$10$^4$ & 1.3\,$\times$10$^4$ & \textbf{1.9\,$\times$10$^4$} \\
\enddata
\tablecomments{(1) ID number of each MMS. (2)-(3) Position of the peak 1.3\,mm flux at each MMS in equatorial coordinates (J2000).
(4) Source size (FWHM) derived from 2D elliptical gaussian fitting. (5) Peak 1.3\,mm flux. (6) Total 1.3\,mm flux.
(7) Velocity width (FWHM) of C$^{18}$O($J=2-1$) emission of MMS-1 and MMS-3 and $^{13}$CO($J=2-1$) emission of MMS-2.
(8) Virial mass calculated from the following relation, $M_{\rm vir}$ = $210 \Delta V^2 R$, where $R$ is the geometric mean of the source size.
(9) Number density of molecular gas derived from the source size and the virial mass.
(10) Mass derived from the total continuum flux, assuming dust temperature of 22\,K \citep{tabatabaei2014}, a gas-to-dust ratio of 300 \citep{relano2018}, and $\kappa_{\rm 1.3~mm}$ of 1\,cm$^2$\,g$^{-1}$ for protostellar envelopes \citep{ossenkopf1994}.
}
\end{deluxetable*}

\section{Discussion} \label{sec:discuss}

\subsection{Triggering mechanism of star formation}

\begin{figure*}[ht!]
\epsscale{1.15}
\plotone{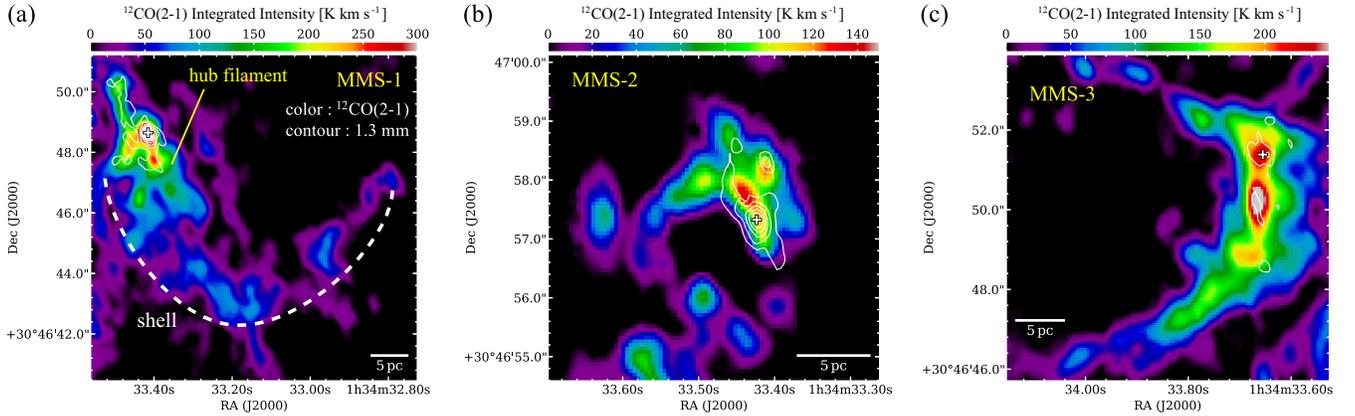}
\caption{Maps of 1.3\,mm continuum flux (contour) superposed on integrated intensity maps in $^{12}$CO($J=2-1$) emission (color) around each molecular shell containing MMSs.
The contour levels are 3, 6, 9, 12, 15, 18 and 21$\sigma$, where 1$\sigma$ = 0.02\,mJy\,beam$^{-1}$. The crosses are the same as those in figure~\ref{fig:12CO13CO}(b).
}\label{fig:12COMMS}
\end{figure*}

Figure~\ref{fig:12COMMS} shows the distribution of molecular gas traced by $^{12}$CO($J=2-1$) emission around each MMS in the NGC~604-GMC.
Every MMS exists within shell-like molecular structures although the radii of molecular shells are different from each other.
In contrast to this, other cloud peaks in $^{12}$CO and $^{13}$CO (figure~\ref{fig:12CO13CO}) do not show such remarkable molecular structures
and also have neither C$^{18}$O nor 1.3\,mm emission tracing dense clumps.
These facts indicate that the molecular clouds associated with the MMSs have experienced a specific compression event to form the massive/dense clumps.

There are some mechanisms which can form such molecular shells and/or trigger high-mass star formation.
Energy inputs from supernovae is a possible candidate.
Supernovae can supply an enormous amount of mechanical energy to surrounding ISM, which is enough to form the molecular shell,
whereas no supernova is found within the three molecular shells \citep[e.g.,][]{garofali2017,long2018}. The role of H$\,${\sc ii} regions containing O-type stars is also important.
For example, strong UV radiation from O-type stars ionize surrounding neutral ISM. Since the ionized region radially spreads, a molecular shell is finally formed.
However, the ionization of neutral ISM cannot induce the high-mass star formation because it does not compress the surrounding molecular gas.
Instead, expansion of H$\,${\sc ii} region is a potential candidate.
Earlier studies (e.g., \citealt{elmegreen1998} and references therein) suggest that
high pressure of H$\,${\sc ii} regions yields direct compression of pre-existing clouds and globules (``globule squeezing''),
and/or gas is accumulated into dense ridge or shell (i.e., the edge of H$\,${\sc ii} regions) and then the cloud collapses gravitationally into dense cores (``collect and collapse'').

\begin{figure*}[htbp]
\epsscale{1.15}
\plotone{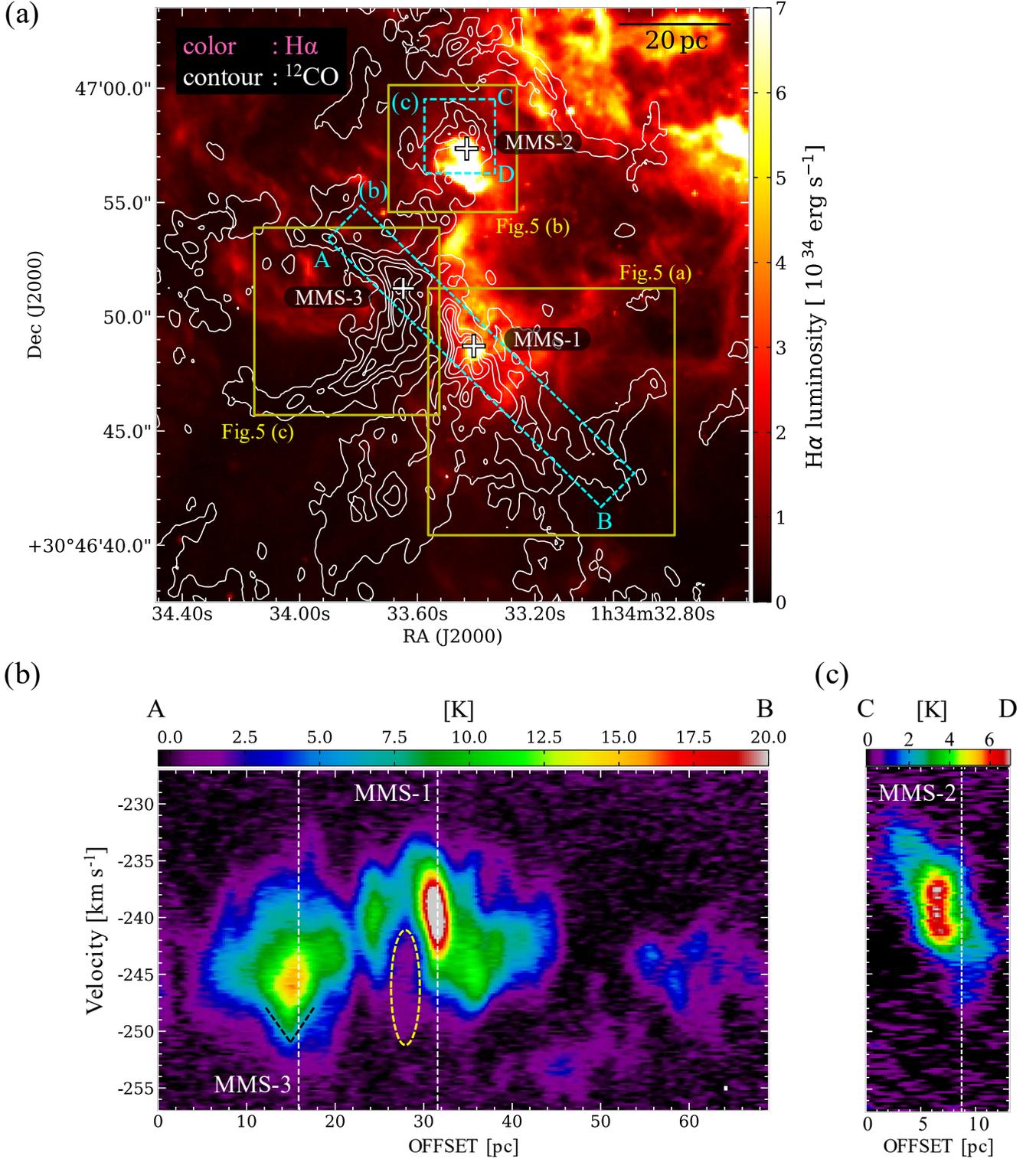}
\caption{(a) Integrated intensity map in $^{12}$CO($J=2-1$) emission (contour) superposed on the map of H$\alpha$ emission (color) around each molecular shell containing MMSs.
The contour levels are 10, 30, 60, 90, 120, 150, 180, and 210\,K\,km\,s$^{-1}$. 
The three yellow rectangles are the same as in figure~\ref{fig:12CO13CO}(a). The crosses are the same as those in figure~\ref{fig:12CO13CO}(b). 
The dashed rectangles show the regions where the position-velocity diagrams are extracted.
(b)--(c) $^{12}$CO($J=2-1$) position-velocity diagrams extracted from the position A to B and from C to D shown in panel (a).
The white vertical lines denote the positions of the MMSs.
The yellow ellipse indicates the possible expansion of the ionized gas which may locally sweep the molecular material.
The black dashed line indicates the high-velocity CO component (called V-shaped feature), suggesting a possible gas colliding signature.
The spatial and velocity resolutions are shown in the lower right corner of panel (b).
}\label{fig:HaPVmap}
\end{figure*}

In figure~\ref{fig:HaPVmap}(a), we examined individual comparison between spatial distributions of the molecular shell and the H$\,${\sc ii} region around each MMS. We found that all the molecular shells with MMSs are associated with H$\,${\sc ii} regions.
In particular, the molecular shell with MMS-2 seems clearly along the edges of their internal H$\,${\sc ii} regions.
A similar spatial correspondence can be observed for the molecular shell with MMS-1 although the H$\alpha$ emission along the western part of the shell is relatively weak.
The molecular shell with MMS-3 is a little away from the edge of the adjacent H$\,${\sc ii} region (in particular, southern part of the shell).
In order to investigate whether the expansion of an individual H$\,${\sc ii} region within each molecular shell triggers high-mass star formation at the MMSs,
we examine the velocity structures based on the $^{12}$CO($J=2-1$) position-velocity (PV) diagrams, which are perpendicularly extracted to the possible expansion direction of each H$\,${\sc ii} region, as shown in figure~\ref{fig:HaPVmap}(b) and (c).
\cite{torii2015} qualitatively described that PV diagrams in CO show an elliptical shape if the expanding motion of the H$\,${\sc ii} region compresses the surrounding molecular materials (see their figure~8).
The PV diagrams observed around the three MMSs show a nearly constant velocity field centered at the systemic velocity of $\sim$240\,km\,s$^{-1}$ with somewhat local fluctuations, rather than a large-scale elliptical shape throughout the clouds.
Note that we found a small elliptical shape at the vicinity of MMS-1 in the PV diagram (figure\,\ref{fig:HaPVmap}(b)).
This implies that the expansion of the ionized gas can {\it locally} sweep the molecular material,
yet it may not contribute to the entire structure formation in their parental molecular clouds and to the high-mass star formation at MMS-1.
This hypothesis is consistent with a numerical study by \cite{shima2017}, suggesting that stellar feedbacks do not primarily affect gas structures in the molecular clouds.
Thus, we concluded that the expansion of H$\,${\sc ii} regions is not likely a trigger of high-mass star formation at MMSs in the NGC~604-GMC.

An alternative potential mechanism which can trigger high-mass star formation is cloud--cloud collision.
In the next subsection, we investigate whether the cloud--cloud collision can explain the high-mass star formation at each MMS.

\subsection{Formation processes of the massive clumps as promising formation sites of high-mass stars}\label{dis:MMS}

As shown in table~\ref{tab:MMS}, the MMSs show massive and dense characteristics. The {\it Spitzer} observations identified bright MIR sources toward the MMSs (subsection \ref{res:1.3mm}), indicating that the high-mass star formation is actively ongoing (see also \citealt{martinez2012}).
Since the three MMSs (and also their parental molecular clouds) are spatially separated as shown in figures~\ref{fig:12CO13CO} and \ref{fig:MMS}, they likely have different origins.
Thus, we discuss possible formation mechanisms of each dense clump, MMS-1, MMS-2, and MMS-3, individually.

\subsubsection{MMS-1} \label{dis:MMS-1}

\begin{figure*}[ht!]
\epsscale{1.15}
\plotone{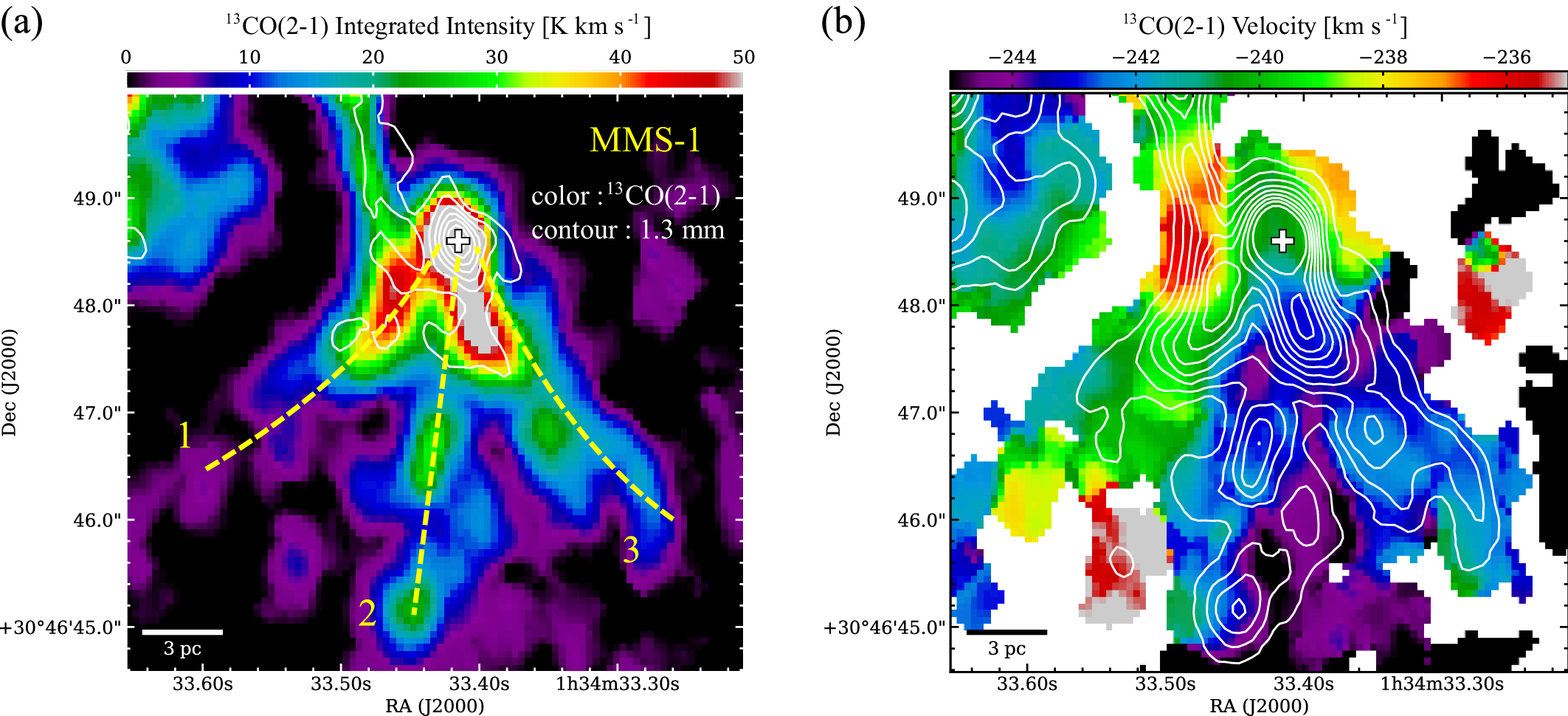}
\caption{(a) Map of 1.3\,mm continuum flux (contour) superposed on integrated intensity map in $^{13}$CO($J=2-1$) emission (color) around MMS-1.
The contour levels are the same as those in figure~\ref{fig:12COMMS}(a). 
The dashed lines indicate individual molecular filaments.
(b) Maps of velocity field (color) and integrated intensity (contour) in $^{13}$CO($J=2-1$) emission around MMS-1.
The contour levels are  6, 12, 18, 24, 30, 36, 42, 48, and 56 K\,km\,s$^{-1}$.
The cross in each panel indicates the peak position of 1.3\,mm continuum flux.
}\label{fig:13COfilament}
\end{figure*}

Figure~\ref{fig:13COfilament}(a) shows the maps of the $^{13}$CO($J=2-1$) integrated intensity and 1.3\,mm continuum emission around MMS-1.
The peak position of the $^{13}$CO($J=2-1$) intensity is almost coincident with not only MMS-1
but also the strong H$\alpha$ spot (see figure~\ref{fig:HaPVmap}(a)), the bright MIR source, and the local peak of 8.4\,GHz continuum emission (see subsection~\ref{res:1.3mm}).
Furthermore, we found that multiple molecular filaments extend from the peak, showing the parasol-like structure.
The 1st moment map (figure~\ref{fig:13COfilament}(b)) shows velocity differences among each filament;
the eastern filament (No.\,1) has a central velocity of $\sim\!-$240\,km\,s$^{-1}$, but those of the other two filaments (Nos.\,2 and 3) are $\sim\!-$243\,km\,s$^{-1}$.
The typical length and mass of the filaments are $\sim$15\,pc and $\sim$3\,$\times$10$^{3}$\,$M_{\odot}$, respectively.
These characteristics are similar to those in molecular filaments associated with 30~--~40\,$M_{\odot}$ protostars in the N159 GMC \citep{fukui2015,saigo2017},
which is one of the most active star-forming regions in the LMC \citep[e.g.,][]{fukui2008,minamidani2008,chen2010}. \cite{fukui2015} and \cite{saigo2017} suggested that filament-filament collisions promoted the high-mass star formation at the intersections.

However, instead of the filament-filament collision scenario, \cite{fukui2019} and \cite{tokuda2019} claimed a large-scale colliding flow scenario in N159 based on higher-resolution studies.
Their $^{13}$CO($J=2-1$) observations found the parasol-like structures composed of more than a few tens filaments, and also found high-mass protostars at the pivots.
Observations toward the Galactic high-mass star-forming regions often identify such complex filamentary configurations, called $``$hub filament$"$\citep[e.g.,][]{myers2009,peretto2013,williams2018}.
Numerical simulations by \cite{inoue2018} suggested that the formation of such a complex/massive filamentary structure, whose line mass exceeds $\sim$1000\,$M_{\odot}$, occurs after a collision between small and large clouds.
\cite{fukui2019} and \cite{tokuda2019} concluded that a large-scale colliding flow is more likely to explain the observed filamentary clouds and the star formation activities therein.
One of the promising candidates of the origin of the large-scale flow is the past tidal interaction between the LMC and the Small Magellanic Cloud \citep{fukui2017,tsuge2019}.

We apply such a mechanism to MMS-1 in NGC~604.
In the case of MMS-1, molecular filaments (i.e., ribs of a parasol) extend southward from the CO peak. This suggests that a small cloud from the south side collided with a pre-existing large cloud. If we assume the trailing spiral arm and gas streaming based on the density-wave theory,
the flowing direction of gas clouds is considered to be from south to north (i.e., upstream to downstream of the spiral arm). 
The speed of the interstellar gas originated from the galactic rotation at NGC~604, $\sim$3.5\,kpc from the galactic center \citep{gratier2012}, is $\sim$100\,km\,s$^{-1}$ \citep{corbelli2014}.
If we subtract a pattern speed, $\Omega_{\rm p}$, of $\sim$25\,km\,s$^{-1}$\,\,kpc$^{-1}$ \citep{newton1980}, the possible colliding gas velocity is likely more than $\sim$10\,km\,s$^{-1}$, which is consistent with the numerical setup by \cite{inoue2018}.
Note that such a large-velocity difference more than $\sim$10\,km\,s$^{-1}$ is $not$ observed around MMS-1 as shown in figure~\ref{fig:13COfilament}(b).
This is because the initial collision velocity is decelerated at the shock front and/or the direction of collision is perpendicular to the line of sight.
The latter effect is especially serious in the case of external galaxies whose inclination is close to face-on. 

In summary, we suggest that the hub filament at MMS-1 was formed by the collision between a small cloud flowing from upstream of the spiral arm and a pre-existing large cloud within the molecular shell, and then high-mass star formation occurred.
\cite{tokuda2020} also found a similar hub-filamentary complex in GMC-16, which is radially extended toward the upstream side of the galactic rotation.
Considering the fact that the strong C$^{18}$O($J=2-1$) emission tracing dense molecular gas is observed at MMS-1, this high-mass star formation seems to be in an earlier stage.

Note that we suppose that the molecular shell with MMS-1 was formed by the ionization due to UV radiation from O-type stars within the internal H$\,${\sc ii} region
because the molecular shell is along the edge of the H$\,${\sc ii} region although the H$\alpha$ emission along the western part of the shell is relatively weak (figure~\ref{fig:HaPVmap}(a)).

\subsubsection{MMS-2 and MMS-3} \label{dis:MMS-2+3}

As shown in figures~\ref{fig:12COMMS}(b) and \ref{fig:HaPVmap}(a), MMS-2 exists at the overlapped region between the strong H$\alpha$ spot of a small H$\,${\sc ii} region and the surrounding molecular shell.
Such a spatial distribution of each emission is quite similar to that of the Galactic H$\,${\sc ii} region RCW~120, which is classified as a $Spitzer$ bubble \citep{churchwell2006}.
According to \cite{torii2015}, the triggering mechanism of high-mass star formation in RCW~120 can be explained by a cloud--cloud collision.
Based on the theoretical model introduced by \cite{habe1992}, they suggested that the exciting O-type star in RCW~120 was formed by a collision between the present two (small and large) clouds at a collision velocity of $\sim$30\,km\,s$^{-1}$.
In this scenario, the molecular shell can be interpreted as the cavity created in the larger cloud by the collision. However, an alternative interpretation is also suggested; for example, \cite{zavagno2020} claimed that they revealed the structure of the dense photodissociation region associated with RCW~120 and evidence for the compression of the internal H$\,${\sc ii} region.
Thus, the formation mechanism of such a ring/shell structure and that of subsequent high-mass stars are just in discussion even for the MW sources.

In the case of MMS-2, the observed velocity structure in $^{12}$CO($J=2-1$) suggest that
the molecular shell is not likely due to the expansion of H$\,${\sc ii} regions as described in the previous subsection.
We thus discuss whether the cloud--cloud collision model can explain the observed molecular shell and high-mass star formation at MMS-2 like RCW~120 \citep{torii2015}.
MMS-2 exists on the western side of the molecular shell. If the cloud--cloud collision model similar to RCW~120 is applied to MMS-2, a small cloud flowed from east to west and finally collided with a large cloud.
This flowing direction of the small cloud is consistent with the infalling H$\,${\sc i} gas suggested by \cite{tachihara2018} (see subsection~\ref{res:CO}).
Therefore, the cloud--cloud collision seems to be an acceptable model to explain the observed feature of MMS-2.

We found no C$^{18}$O($J=2-1$) emission at MMS-2 as described in subsection~\ref{res:1.3mm}.
However, the molecular gas density at MMS-2 exceeds 10$^4$\,cm$^{-3}$ (see table~1), which is comparable to those at MMS-1 and MMS-3 where the significant C$^{18}$O($J=2-1$) emission is detected.
This is likely explained by the selective photodissociation of C$^{18}$O molecule due to the far-UV radiation as in the cases of the Galactic molecular clouds \citep[e.g.,][]{lada1994,buckle2012,shimajiri2014}.
We thus consider that the high-mass star formation at MMS-2 is in a later stage than the other MMSs.

Then we explore whether the cloud--cloud collision model can explain MMS-3 and its surrounding molecular shell.
Since MMS-3 exists in the western side of the molecular shell similar to the case of MMS-2, the infalling H$\,${\sc i} gas likely contributes to the formation of the molecular shell through the cloud--cloud collision.
In fact, we found a possible gas colliding signature on the PV diagram (figure~\ref{fig:HaPVmap}(b)); there exists a high-velocity component, called V-shaped feature (e.g., \citealt{fukui2018}), centered at the position of MMS-3.
However, the H$\alpha$ emission is relatively weak inside the molecular shell with MMS-3 compared to that with MMS-2.
According to the extinction map of the NGC~604-GMC \citep[e.g.,][]{maiz2004,relano2009}, extinction at MMS-3 is comparable to that at MMS-2.
Thus, the weak H$\alpha$ emission at MMS-3 is not caused by the strong extinction by dust, but presumably due to the inactive high-mass star formation.

A similar case has been observed toward the Galactic source;
\cite{higuchi2014} found a molecular shell possibly formed by a cloud--cloud collision at G0.253+0.016 (G0.25 cloud) in the Galactic center,
whereas the high-mass star formation is inactive inside the shell.
They concluded that the G0.25 cloud is in the very early stage of star formation, triggered by the cloud--cloud collision.
We thus suppose that the high-mass star formation at MMS-3 is in an earlier stage, which is consistent with the existence of dense molecular gas traced by the moderate C$^{18}$O($J=2-1$) emission.

\subsubsection{A possible origin of different morphologies in the molecular clouds around the MMSs} \label{dis:MMS-summary}

The parental clouds of MMS-2 and MMS-3 show shell-like structures, whereas that of MMS-1 is the hub filament composed of at least three components.
The former is similar to collision products by the Habe \& Ohta model \citep{habe1992,takahira2014,shima2018}.
The latter can be reproduced by \cite{inoue2018} model as discussed in section~\ref{dis:MMS-1}.
This discrepancy is presumably due to the differences in the initial conditions of the cloud--cloud collisions between the two models.
In the Habe \& Ohta model, they simulated a collision between two spherical clouds,
while \cite{inoue2018} treated a collision between a small spherical cloud and a large cloud surface filling a part of the numerical cubic box.

Magnetic field properties may be also important because dense filamentary structures are efficiently formed when the magnetic field orientation is perpendicular to the collision direction \citep{inoue2018}.
Recently, \cite{sakre2020} investigated the effect of magnetic field on the Habe \& Ohta model.
When the collision direction is perpendicular to the strong (4\,$\mu$G) magnetic field, the collision product is close to
our detected $``$hub$"$ composed of three linear-shaped filaments rather than a shell-structure, which is formed in the parallel magnetic field case.
From an observational perspective, a radio-continuum polarization study by \cite{tabatabaei2008} revealed that the magnetic field orientation toward the NGC~604-GMC is in the east-west direction.
This might explain the observed molecular-gas structures around MMSs;
i.e., the cloud--cloud collision in the north-south direction (i.e., perpendicular to the magnetic field) at MMS-1 formed the hub-filamentary shape,
whereas those in the east-west direction (i.e., parallel to the magnetic field) at MMS-2 and MMS-3 produced shell-like features.
However, the actual magnetic field properties toward the molecular phase are still unknown because the available CO polarization measurement with a much higher spatial resolution ($\sim$15\,pc) toward M33 by \cite{li2011} intrinsically contains the 90$\arcdeg$ ambiguity of the Goldreich-Kylafis effect.
Future high-resolution/sensitivity polarization observations combining CO and thermal dust emission will complementarily constrain the magnetic field properties in the NGC~604-GMC.

\subsection{Star formation scenario in NGC~604}

For NGC~604, a scenario of sequential star formation has been discussed.
In particular, \cite{tosaki2007} found a high CO($J=3-2$)/CO($J=1-0$) ratio gas arc along the bright H$\alpha$ nebula,
and suggested that the high-ratio gas arc revealed the presence of warm and dense molecular gas layers, which were compressed by stellar wind and/or supernovae from the central star cluster of NGC~604. They concluded that the arc-like structure of the high-ratio gas is the site of second-generation star formation triggered by first-generation star formation (i.e., the central star cluster).

The location of the high-ratio gas arc roughly corresponds to the northern field of the ALMA observations, where strong $^{12}$CO($J=2-1$) and $^{13}$CO($J=2-1$) emission and 1.3\,mm continuum emission are observed (see figures~\ref{fig:12CO13CO} and \ref{fig:MMS}).
Thus, the three MMSs are considered to be at least a part of the second-generation star formation suggested by \cite{tosaki2007}.
However, the triggering mechanism of the high-mass star formation at the MMSs is not likely the feedback from the central star cluster and/or H$\,${\sc ii} regions, but cloud--cloud collisions as discussed in the subsection~\ref{dis:MMS}.
We suppose that there is little relationship between the ongoing high-mass star formation at MMSs and the central star cluster of NGC~604.
Instead, the galactic-scale gas dynamics plays an important role in the ongoing high-mass star formation in NGC~604.

We compare our results with numerical simulations of isolated spiral galaxies.
\cite{dobbs2018} found that a model with higher feedback best reproduced the M33-like spiral structure in the gas.
The higher feedback generally makes neighboring molecular clouds smaller.
In fact, small clouds whose typical size is less than 10\,pc are observed in the southern part of the NGC~604-GMC (see figure~\ref{fig:12CO13CO}).
Such a small cloud may produce the observed hub filament at MMS-1 through the cloud-cloud collision.
\cite{dobbs2015} suggested that moderate- and long-lived clouds experience one or more cloud-cloud collisions within their lifetime.

\begin{figure*}[ht!]
\epsscale{1.15}
\plotone{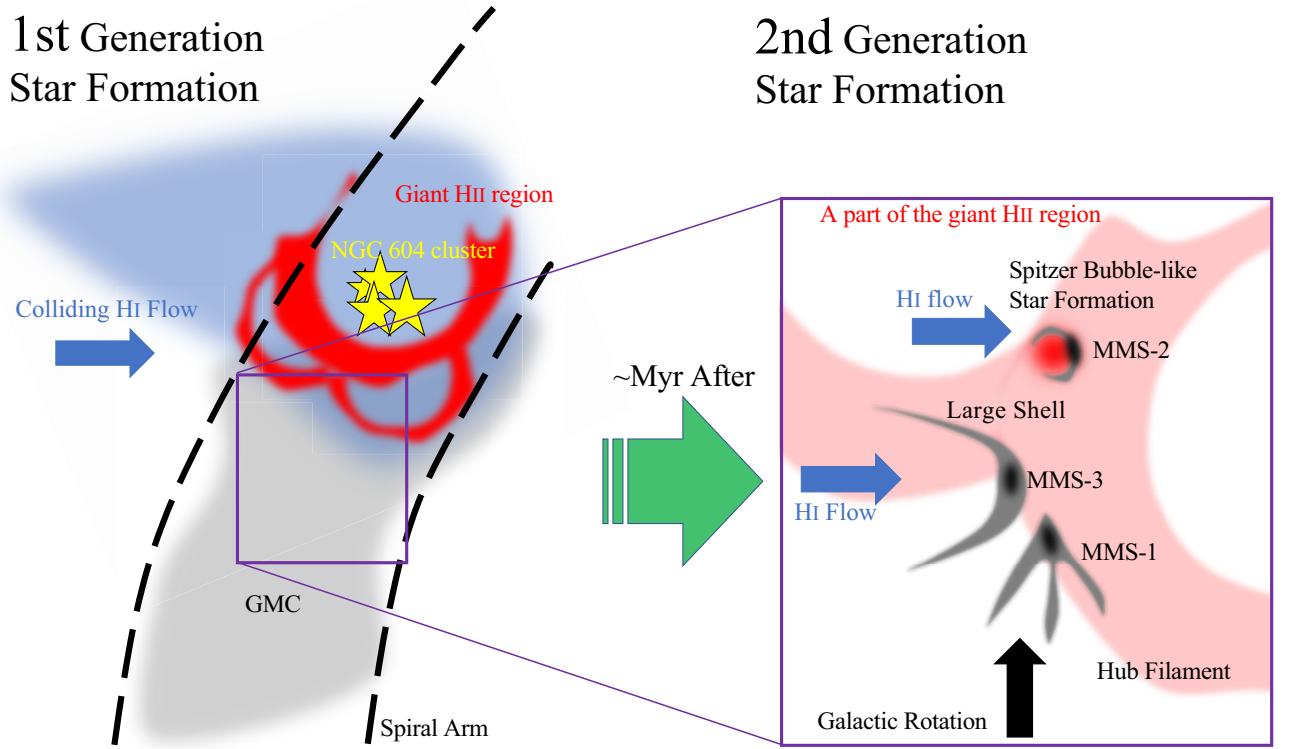}
\caption{A schematic view of star formation process in NGC~604 motivated by the ALMA observations. Note that the colliding H$\,${\sc i} gas was proposed by \cite{tachihara2018}
}\label{fig:Cartoon}
\end{figure*}

In figure~\ref{fig:Cartoon}, we summarized the schematic view of our proposed star formation scenario in NGC~604.
\cite{tachihara2018} found two H$\,${\sc i} components with a velocity separation of $\sim$20\,km\,s$^{-1}$ in NGC~604, and suggested that the blueshifted velocity components triggered the first-generation star formation.
Even after the drastic cluster formation, the converging gas is still energetic enough to propagate the star formation activities southward.
The colliding flow remnant triggered the formation of MMS-2, which is likely the most evolved one among the three MMSs, and subsequently formed MMS-3 and its surrounding molecular structure.
Note that the parental molecular gas of MMS-1 is possibly originated from the pre-existing cloud in the southern side of NGC~604; i.e., its origin is $not$ converging H$\,${\sc i} gas flow, but likely the galactic rotation.
The role of the galactic rotation in the compression of the molecular material has been already pointed out by \cite{tosaki2007}.

The NGC~604 complex contains one of the most mass-concentrated GMC in M33,
and the molecular gas structures show well-developed filaments, which have dense clumps identified at the millimeter wavelength.
In contrast to this, GMC-16 shows filamentary clouds with a length of $\sim$50\,pc along with the spiral arm \citep{tokuda2020}
and GMC-8 shows an extended round-shaped structure without any remarkable dense filaments/clumps (H. Kondo et al. in preparation)
although the two GMCs have a comparable molecular mass ($\sim 10^6$\,$M_{\odot}$) to the NGC~604-GMC.
The resolved GMC structures may provide us crucial hints to understand their formation history.
In the case of NGC~604, we suggest that galactic-scale gas flows from at least two directions work to build up the largest high-mass star-forming complex in the Local Group of galaxies.

\section{Summary}

We have performed ALMA observation in $^{12}$CO($J=2-1$), $^{13}$CO($J=2-1$), and C$^{18}$O($J=2-1$) lines and 1.3\,mm continuum emission
toward a GMC associated with the giant H$\,${\sc ii} region NGC~604 at an angular resolution of 0\farcs44 $\times$ 0\farcs27 (1.8\,pc $\times$ 1.1\,pc).
The summary of this work is as follows:

\begin{enumerate}
\item
The molecular gas distributions in $^{12}$CO and $^{13}$CO lines are highly structured,
i.e., composed of a lot of molecular filaments, arcs, and shells whose lengths are 5~--~20\,pc, indicating an evolved nature of the GMC.
The total H$_2$ gas mass estimated from the $^{12}$CO and $^{13}$CO data are $\sim$3\,$\times$10$^{6}$\,$M_{\odot}$ and $\sim$3\,$\times$10$^{5}$\,$M_{\odot}$, respectively.

\item
We have identified three 1.3\,mm continuum sources (MMS-1, MMS-2, and MMS-3) with a gas mass of $\sim$10$^4$\,$M_{\odot}$ at edges of the hub filament and shells.
We detected C$^{18}$O emission at MMS-1 and MMS-3 but not in MMS-2. This discrepancy is presumably due to their different evolutionary stages.
The presence of the massive/dense clumps strongly indicates that the high-mass star formation is still ongoing even after the formation of the first-generation cluster embedded in the giant H$\,${\sc ii} region.

\item
We revealed that the $^{12}$CO peak temperature at the northern molecular complex is as high as $\sim$40\,K, and some of the molecular shells are along the edge of H$\,${\sc ii} regions.
These observational facts indicate that the giant H$\,${\sc ii} regions contribute to the heating and the ionization of surrounding ISM.
However, the CO velocity structures indicate that the expanding motion of H$\,${\sc ii} regions does not promote the compression of the molecular material to trigger the second-generation star formation.
Alternatively, multiple cloud--cloud collisions induced by an external H$\,${\sc i} gas flow and the galactic rotation may explain
the formation of the molecular shells and the hub filament associated with dense clumps as the birth sites of high-mass stars.

\end{enumerate}

\acknowledgments
We thank the anonymous referee for helpful comments, which significantly improved the manuscript.
This paper makes use of the following ALMA data: \dataset[ADS/JAO.ALMA\#2017.1.00461.S]{https://almascience.nrao.edu/aq/?project_code=2013.1.00212.S}
ALMA is a partnership of ESO (representing its member states), NSF (USA) and NINS (Japan), together with NRC (Canada),
NSC and ASIAA (Taiwan), and KASI (Republic of Korea), in cooperation with the Republic of Chile.
The Joint ALMA Observatory is operated by ESO, AUI/NRAO and NAOJ.
This work is based on observations made with the NASA/ESA Hubble Space Telescope,
obtained from the data archive at the Space Telescope Science Institute.
STScI is operated by the Association of Universities for Research in Astronomy, Inc.\,under NASA contract NAS 5-26555.
This work was supported by NAOJ ALMA Scientific Research grant Nos. 2016-03B and JSPS KAKENHI (grant Nos. 17K14251, 18K13580, 18K13582, 18K13587, 18H05440, 19K14758, and 19H05075).
\software{CASA (v5.4.0; \citealt{mcmullin2007}), Astropy \citep{astropy2018}, APLpy (v1.1.1; \citealt{Robi12})}

\end{document}